\documentclass[final,3p,times,twocolumn]{elsarticle}

\usepackage{amssymb}
\usepackage{amsmath}
\usepackage{url}
\usepackage{hyperref}

\usepackage{breakurl}

\def \figpath {figures/}

\journal{Fusion Engineering and Design}

\begin{document}
\begin{frontmatter}


\title{Overview of the Helios Design: A Practical Planar Coil Stellarator Fusion Power Plant}

\author[Thea]{C.P.S. Swanson} 
\author[Thea]{S.T.A. Kumar}

\author[Thea]{D.W. Dudt}
\author[Thea]{E.R. Flom}
\author[Thea]{W.B. Kalb}
\author[Thea]{T.G. Kruger}
\author[Thea]{M.F. Martin}
\author[Thea]{J.R. Olatunji}
\author[Thea]{S. Pasmann}
\author[Thea]{L.Z. Tang}
\author[Thea]{J. von der Linden}
\author[Thea]{J. Wasserman}

\author[Thea]{M. Avida}
\author[Thea]{A.S. Basurto}
\author[Thea]{M. Dickerson}
\author[Thea]{N. de Boer}
\author[Thea]{M.J. Donovan}
\author[Thea]{A.H. Doudna Cate}
\author[Thea]{D. Fort}
\author[Thea]{W. Harris}
\author[Thea]{U. Khera}
\author[Thea]{A. Koen}
\author[Columbia]{J.A. Labbate}
\author[Thea]{N. Maitra}
\author[Thea]{A. Ottaviano}
\author[Thea]{R.K. Parmar}
\author[Columbia]{E.J. Paul}
\author[Thea]{B. Reydel}
\author[Thea]{A. van Riel}
\author[ind]{P.K. Romano}
\author[Thea]{M. Savastianov}
\author[Thea]{S. Saxena}
\author[Thea]{S. Seethalla}
\author[Thea]{S. Srinivasan}
\author[Thea]{R.H. Wu}

\author[Thea]{D. Nash}
\author[Thea]{J. Priebe}
\author[Thea]{M. Slepchenkov}
\author[Thea]{S. Walsh}

\author[Thea]{B. Berzin}
\author[Thea]{D.A. Gates}
\author{and the Thea Energy team}
\affiliation[Thea]{organization={Thea Energy, Inc.},
            addressline={1 Eastern Road, Suite 3-04},
            city={Kearny},
            postcode={07032},
            state={NJ},
            country={USA}}
\affiliation[Columbia]{organization={Department of Applied Physics and Applied Mathematics, Columbia University},
            city={New York},
            state={NY},
            postcode={10027},
            country={USA}}
\affiliation[ind]{organization={Independent contractor}}

\begin{abstract}
Thea Energy, Inc. has developed the preconceptual design for ``Helios,'' a fusion power plant based on the planar coil stellarator architecture. In this overview paper, the design is summarized and the reader is referred to the papers in this special issue for more detail. The Helios design is based around a two-field-period quasi-axisymmetric (``QA'') stellarator equilibrium with aspect ratio 4.5 and a novel tokamak-like X-point divertor. The natural stability, low recirculating power, and steady-state capability of the stellarator are leveraged. Stability and transport are calculated using state-of-the-art, high-fidelity codes and grounded in measured performance of existing experiments. The electromagnetic coil set is high-temperature superconducting (``HTS'') and consists of 12 large, plasma-encircling coils like the toroidal field coils of a tokamak, and 324 smaller, field-shaping coils. All coils are planar and convex. A maximum of 20 T on-coil is enforced, a value which has been achieved in existing large-bore HTS coils. There is a minimum of 1.2 m between plasma and coils, leaving space for tritium breeding blanket and neutron shielding. Because of this thick shielding, all coils have a minimum 40-year operational lifetime, the same minimum lifetime of the overall power plant system. 1.1 GW of thermal power and 390 MW of net electric power are produced. The shaping coils are individually controllable, enabling a uniquely configurable magnetic field for relaxed manufacturing and assembly tolerances as well as plasma control in the presence of bootstrap current. A practical maintenance architecture is a primary driver of the design; maintenance is performed on entire toroidal sectors that are removed from between the encircling coils. A biennial maintenance cycle is estimated to take approximately 84 days, resulting in an 88\% capacity factor. In all systems, rigorous engineering constraints such as temperature and stress limits are enforced. 

\end{abstract}

\end{frontmatter}

\section{Introduction}
\label{sec:intro}

Thea Energy, Inc. has developed the preconceptual design for ``Helios,'' a fusion power plant based on the planar coil stellarator architecture \cite{gates_planar_2024,gates_stellarator_2025,swanson_scoping_2025,kruger_coil_2025,wu_planar_2025}. In this overview paper, the design is summarized and the reader is referred to the papers in this special issue for more detail \cite{kumar_heating_nodate,dudt_equilibrium_nodate,von_der_linden_alpha_nodate,martin_mhd_nodate,martin_profile_nodate,kruger_planar_nodate,flom_design_nodate,olatunji_design_nodate,olatunji_design_nodate-1,kalb_preliminary_nodate,pasmann_2025_preliminary,tang_facility_nodate,wasserman_helios_nodate,slepchenkov_electrical_nodate,slepchenkov_control_nodate}.

As a plasma confinement device, the stellarator \cite{spitzer_stellarator_1958,helander_theory_2014,imbert-gerard_introduction_2019} has a mature physics basis, alongside the tokamak \cite{helander_stellarator_2012}. There are more than a dozen large ($R>1$ m), high-field ($B_0>1$ T) stellarator experiments in the International Stellarator Database \cite{yamada_characterization_2005}, spanning four decades and many countries. Critical performance parameters such as stability and transport may now be predicted with reasonable accuracy. 

From a power plant basis, stellarators are recognized as having practical advantages over other architectures. There is no need to drive electrical current within the plasma itself, leading to an inherently steady-state and low-recirculating-power facility. The stability of the stellarator is excellent, and they are immune from the damaging disruptions seen in tokamaks. Finally, there is no Greenwald density limit in stellarators, permitting them to operate at higher density than tokamaks \cite{peterson_density_2005,gates_origin_2012}. Comparative analyses of tokamak-based and stellarator-based pilot plant architectures have favored the latter \cite{menard_prospects_2011}.

However, stellarators initially lagged behind tokamaks in this area. The earliest power plant design studies used what would now be called un-optimized stellarators such as the torsatron and the heliac \cite{lyon_status_1994,lyon_physics_1994}. Consequently these designs were predicted to lose significant fractions ($> 40\%$) of the fusion-product alpha particles. These designs had to make compromises to produce significant self-heating. Some were impractically large. Some were designed to operate colder and denser than optimal, resulting in a lower power density. Some had to extend plasma physics assumptions into the realm of the implausibly aggressive. 

A turning point in the field of stellarator theory came in the 1980s and 1990s, when categories of stellarators were developed whose particle transport were similar to that of tokamaks, so-called optimized stellarators \cite{pytte_neoclassical_1981,boozer_transport_1983,nuhrenberg_stable_1986,garren_magnetic_1991,helander_theory_2014}. Early power plant designs based on these configurations revealed that the improved thermal and energetic particle confinement was enabling to the concept \cite{lyon_physics_1994,miller_stellarator_1997}, though new difficulties were identified (an alternate research track continues on the Heliotron approach, informed by the LHD experiment \cite{takeiri_prospect_2018,goto_importance_2011,miyazawa_development_2023}). These optimized stellarators required so-called modular coils, 3D-curved electromagnetic coils with no particular symmetry and tight hardware tolerances. In these optimized stellarators, close proximity between the coils and the plasma boundary was required. This latter requirement resulted in designs which were often just as large as prior generations of un-optimized stellarators. \cite{lyon_physics_1994,warmer_system_2016}.

At the same time, three modular-coil experiments were designed to implement these optimized stellarators: HSX, NCSX, and W7-X \cite{anderson_helically_1995,zarnstorff_physics_2001,beidler_physics_1990}. Of these, all three exhibited cost and schedule overruns, and identified significant practical difficulty in designing, manufacturing, and assembling complexly curved, 3D coils to the required precision \cite{geiger_hsx_2024,chrzanowski_lessons_2009,neilson_lessons_2009,bosch_engineering_2018}. NCSX was canceled partway through manufacturing. 

Recent breakthroughs led to the possibility of a practical, compact stellarator power plant. The first was the development of quasi-axisymmetric (``QA'') stellarator equilibria. This type of equilibrium can be more compact, their boundaries can be less strongly magnetically shaped, their coils can consequently be further away from the boundary \cite{zarnstorff_physics_2001,ku_physics_2008,kappel_magnetic_2024}. The second was the invention of the planar coil stellarator architecture, which utilizes planar, convex coils that can be manufactured conventionally by winding in tension, and individually controllable planar field-shaping coils that can both correct for hardware defects or assembly errors and control the bootstrap current of the QA plasma \cite{gates_planar_2024,gates_stellarator_2025,swanson_scoping_2025,kruger_coil_2025,wu_planar_2025}. The third was the commercialization of high-temperature superconductor (``HTS'') \cite{grant_cost_2002,mitchell_superconductors_2021,molodyk_prospects_2023}, which can carry significantly higher current, at a higher magnetic field, at a higher temperature, than Low Temperature Superconductor (``LTS''). This higher current density and temperature is enabling to the compactness and power balance of the superconducting stellarator. 

Combined here for the first time, the Helios power plant design leverages all three of these features. The design considers the interrelation between physics, engineering, and economic considerations. Helios does not have the highest power density of the realistically proposed fusion power plants; nor does it have the highest magnetic field or beta; nor is it the most compact. Rather, it combines engineering constraints that are either known (temperature, stress) or extrapolated (neutron damage thresholds, heat flux) with achieved normalized plasma performance (confinement, stability), into an integrated power plant design. Practicality, conservatism, and engineering margin are primary design drivers. 

\begin{figure}
\centering
\includegraphics[width=1.0\linewidth]{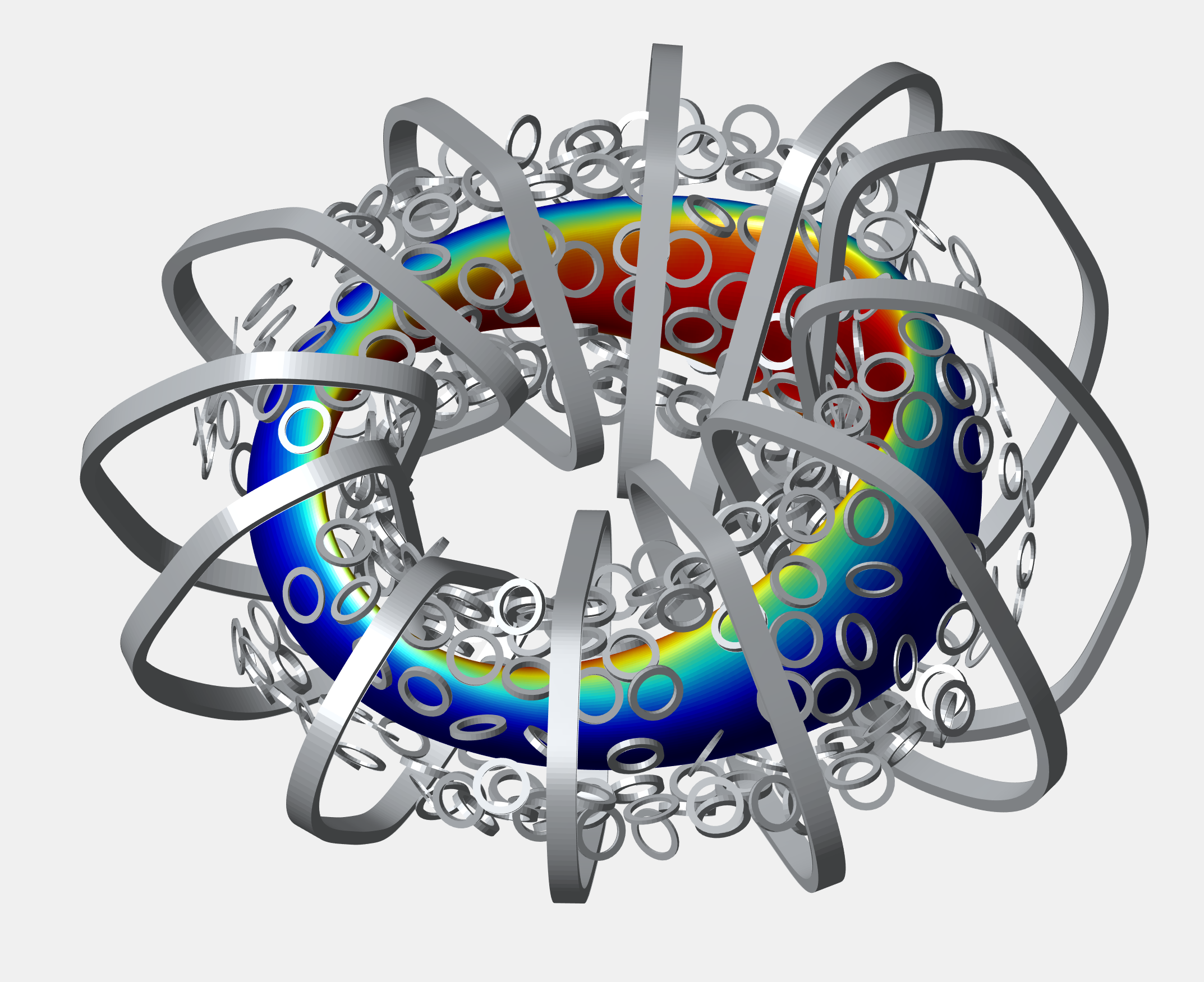}
\caption{The Helios equilibrium and coil set. The coils (silver) are planar and convex. The quasi-axisymmetric plasma equilibrium (red to blue) is shaded by the magnetic field amplitude on the surface produced by that coil set. Very good quasi-symmetry is visible.}\label{fig:coils_eq}
\end{figure}

An important touchstone in the field of stellarator power plant design is the ARIES-CS study \cite{najmabadi_aries-cs_2008,ku_physics_2008}. Helios bears a resemblance to that design. The size and magnetic field strength are similar. However, the assumed plasma beta and confinement multiplier are significantly less aggressive. These changes are informed by higher-fidelity modeling than what was available when ARIES-CS was designed, as discussed later in this paper. 

Additionally, the quasi-symmetry error of the Helios equilibrium is lower than that of ARIES-CS, owing in part to advances in the optimization procedure \cite{landreman_magnetic_2022}. This results in improved energetic particle confinement, easing self-heating thresholds and decreasing the constraints on the divertor. This degree of quasi-symmetry can be seen in the smooth, toroidal bands of $|B|$ visible on the plasma boundary shown in Figure \ref{fig:coils_eq}. 

The Helios equilibrium is less strongly shaped, allowing the coils to be further away. This significantly eases the design of the breeding blanket while protecting the coils for the 40-year lifetime of the plant. ARIES-CS required a highly optimized non-uniform blanket that prioritizes shielding at the expense of breeding zone in certain places \cite{el-guebaly_designing_2008}.

The ARIES-CS coil structure was 3,000 tons and envisioned to be 3D printed on-site, as it was too big to transport \cite{waganer_aries-cs_2008b}. Tight coil tolerances were required, perhaps impractically so. The Helios coils, on the other hand, are all planar and convex, and can be wound in tension. The coil winding packs can be seen in Figure \ref{fig:coils_eq}. Additionally, tolerances are significantly relaxed as manufacturing and assembly errors can be corrected during operation by the device's control system, which independently adjusts the operating currents of the shaping coils.

Because of the small gaps between ARIES-CS coils, there were only three small ports envisioned for maintenance. Consequently 222 individual components had to be serially removed through these ports \cite{waganer_aries-cs_2008}. Helios has large gaps between large, planar encircling coils, and the shaping coils can be removed from between them. Entire toroidal sectors may be removed and replaced at a time.

At one point during the ARIES-CS development, the field-period-based maintenance scheme was considered \cite{wang_maintenance_2005}. In this scheme, entire field periods are removed and maintained. ARIES-CS rejected this approach because it involved removing 4,000-ton components (entire thirds of the stellarator), and the removal, replacement, and realignment of the superconducting coils each time. The Helios sector-based maintenance scheme removes much smaller components, toroidal sectors that can fit between the encircling coils. The encircling coils do not have to be removed and re-integrated. 

Together, these considerations result in a design which is significantly more practical than a plant based on an approach similar to ARIES-CS. 

\section{Summary of the design}
\label{sec:summary}

\begin{figure}
\centering
\includegraphics[width=1.0\linewidth]{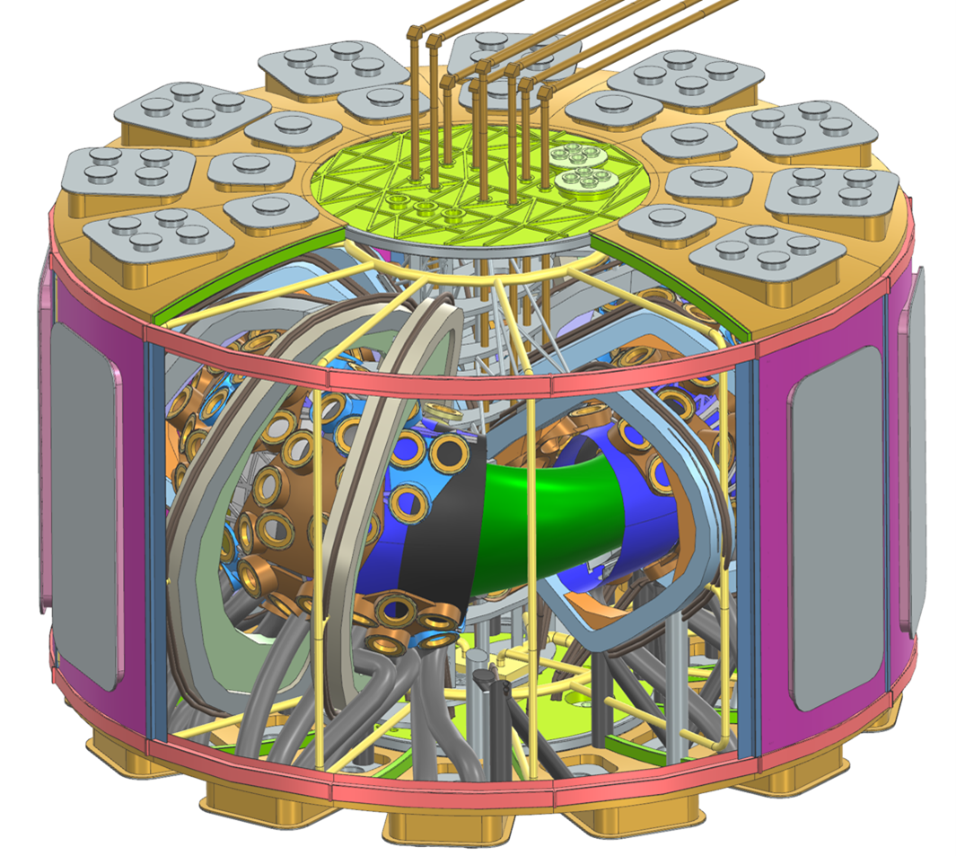}
\caption{Helios architecture, cut-away view. From interior to exterior, the plasma (green), the blanket (black and blue), the shaping coils (orange), the encircling coils (gray). The cryostat surrounds all with removable panels. The divertor pump ducts, cryogen delivery manifold, and microwave waveguides are also shown.}\label{fig:Cutaway}
\end{figure}

The most important global parameters of the Helios facility are tabulated in Table \ref{tab:parameters}. A rendering of the stellarator itself, including many of the key features, can be seen in Figure \ref{fig:Cutaway}. 

Helios has an 8 m major radius, aspect ratio 4.5, and 6 T axial magnetic field. It is a two-field-period QA stellarator. Its plasma is a mixture of deuterium and tritium and undergoes thermonuclear fusion at a rate sufficient to produce 960 MW of fusion power. The plasma is started up by high-frequency microwaves (electron cyclotron resonance heating) and fueled by gas puffing at the edge and pellet fueling of the core. Operating in steady state it is essentially ignited, self-heating via the fusion reaction. The stability and confinement of the Helios plasma was targeted using the normalized performance of existing stellarators and verified using high-fidelity, 3D models of the relevant phenomena. 

\begin{table*}
\centering
\begin{tabular}{||l|c|c||} 
\hline
\textbf{Parameter} & \textbf{Symbol} & \textbf{Quantity} \\ 
\hline\hline
Major radius & $R$ & 8 m \\ \hline
Aspect ratio & $A$ & 4.5 \\ \hline
Minor radius & $a$ & 1.8 m \\ \hline
Magnetic field on-axis & $B_0$ & 6 T \\ \hline
Auxiliary ECRH heating power & $P_{aux}$ & 
      \begin{tabular}{c} 10 MW (startup) \\ <1 MW (ignited) 
      \end{tabular} \\ \hline
Volume-averaged beta & $\langle \beta\rangle_V$ & 2.7 \% \\ \hline
Rotational transform at 2/3 surface & $\iota_{2/3}$ & 0.46 \\ \hline
ISS04 confinement enhancement factor & $H_{ISS04}$ & 1.4 \\ \hline
Energy confinement time & $\tau_E$ & 1.8 s \\ \hline
Peak electron density & $n_{e0}$ & $2.1\times10^{20}$ /m$^3$ \\ \hline
Peak ion temperature & $T_{i0}$ & 20 keV \\ \hline
Sudo density limit multiplication factor & $f_{Sudo}$ & 
      \begin{tabular}{c} 1.25 (startup) \\ 1.1 (ignited)
      \end{tabular} \\ \hline 
Plasma volume & $V$ & 500 m$^3$ \\ \hline
Fusion power & $P_{fus}$ & 958 MW \\ \hline
Total thermal power & $P_{therm}$ & 1.1 GW \\ \hline
Thermal conversion efficiency & $\eta$ & 40 \%  \\ \hline
Net electric power & $P_{net,e}$ & 390 MW$_e$  \\ \hline
Magnet operating temperature & $T_{coil}$ & 20 K \\ \hline
Maximum magnetic field on-coil & $B_{max}$ & 20 T \\ \hline
Minimum plasma-coil distance &  & 1.2 m \\ \hline
Idealized tritium breeding ratio & $TBR$ & 1.3 \\ \hline
Coil minimum lifetime &  & 40 years \\ \hline
Tritium startup inventory & $I_0$ & 1-2 kg \\ \hline
\hline
\end{tabular}
\caption{A Summary of Helios's key geometric, physics, and engineering properties}
\label{tab:parameters}
\end{table*}

Helios has a novel toroidally continuous non-resonant X-point divertor like that of a tokamak, a first for a stellarator power plant design. This divertor can be expected to exhaust gas 10 times more effectively than existing stellarator divertors \cite{feng_comparison_2011}. The Helios design incorporates a tokamak-like divertor into a fully optimized stellarator configuration, leveraging decades of practical tokamak experience and permitting more conservative vacuum-pumping solutions. The divertor targets are tungsten, cooled with helium. 

Helios is designed with a minimum of 1.2 m between the plasma and any part of a coil. This permits a uniform radial build. The first wall is a vanadium alloy, chosen for its long survival (15 years) under high-energy neutron flux. The tritium breeding blanket is a lead-lithium eutectic with 65\% isotopic enrichment of lithium-6. The idealized tritium breeding ratio is 1.3. There is ample room for a multi-layer neutron shield, which limits the heating and neutron damage of the coils. The blanket, first wall, and shield are cooled with helium. The coils are well-shielded from neutrons and last a minimum 40-year lifetime (the lifetime of the stellarator), a key enabler of economic operation. An approximately two-meter-thick concrete bioshield surrounds the stellarator hall. The neutronic properties were modeled using 3D Monte Carlo neutron and photon transport simulation. 

Including the tritium breeding reaction in the blanket, 1.1 GW of thermal power is produced in total. 390 MW of net electric power is produced via a steam Rankine cycle. Around 1-2 kg of tritium is required to start up the plant; thereafter it is self-sufficient with respect to tritium. 

The coils are HTS operated at 20 K. All coils are planar and convex. 12 plasma-encircling coils act similar to tokamak toroidal field coils. The 324 field-shaping coils are individually controllable, permitting good quasi-symmetry and divertor strike point control during startup and through to ignited operation. Coil tolerances can be relaxed by magnetic configurability of this type. The encircling coils are insulated. Upon a detected quench, their currents are actively dumped into external resistors. The shaping coils are partially-insulated and self-protecting in the case of a quench.

Helios is designed subject to the constraint that the maximum field on-coil is 20 T. This value is set via a tradeoff between magnet practicality and fusion power density. Large-bore high-field HTS magnets have achieved 20.1 T in practice \cite{hartwig_sparc_2023}. Using achieved physics and engineering limits such as this field limit is deemed an engineering requirement. If, in the future, 25 T on-tape is found to be plausible, the field on-axis would be 7.5 T, the total fusion power would be 2.3 GW and the net electric power would be 1.0 GW. However, a higher-field coil is more likely to destructively quench, with 60\% higher quench energy released at 25 T. Likewise, the magnetic forces are 60\% higher. Using a 20 T maximum field permits the material stresses to remain within the capabilities of ordinary steel and not require the use of exotic alloys. The quench dynamics of the coil are modeled using multi-physics COMSOL. The coil support structure is modeled using a commercial 3D CAD/FEA package. 

The Helios maintenance operation occurs during one planned outage of approximately 84 days every two years, enabling an 88\% capacity factor. Entire toroidal sectors of the radial build are removed from between the encircling coils. 

\section{Plasma design and simulation}
\label{sec:plasma}

In this section, the design of the Helios equilibrium, plasma, operational scenario, and divertor is discussed in more detail. Each subsection discusses a plasma physics phenomenon. The reader is referred to the other papers in this special issue for more detail on each subject. 

In Section \ref{sec:scoping}, a low-fidelity scoping activity is discussed. In Section \ref{sec:equilibrium}, the plasma equilibrium is described. In Section \ref{sec:alphas}, simulations of energetic particle confinement are shown. Section \ref{sec:MHD} discusses considerations of magnetohydrodynamic stability and evolution. In Section \ref{sec:turbulence}, turbulent transport is simulated, and the temperature profile evolution is simulated under self-consistent transport models. In Section \ref{sec:coils_phys}, the physics design of the coils is provided. In Section \ref{sec:divertor_phys}, the tokamak-like X-point divertor is described and shown. 

\subsection{Scoping studies, heating and fueling, and dynamic accessibility}
\label{sec:scoping}
Before detailed equilibrium design and plasma physics analysis, reduced 0D scoping models were used to target the stellarator scale and bulk parameters. 1D profile-dependent models were further used to refine the design and develop operational and startup scenarios. 

The 0D scoping step is common and commensurate with other stellarator systems codes \cite{swanson_scoping_2025,lyon_systems_2008,lyon_status_1994,lion_general_2021,goto_importance_2011}. Transport in Helios is assumed to follow the ISS04 scaling \cite{yamada_characterization_2005}, with a confinement enhancement multiplier of $H_{ISS04} = 1.4$. This value has been achieved in the W7-X stellarator \cite{beurskens_ion_2021,bozhenkov_high-performance_2020}. This assumption is verified by self-consistent gyrokinetic calculations in Section \ref{sec:turbulence}. The empirical Sudo line-averaged density limit is used \cite{sudo_scalings_1990}, and is exceeded by less than 10\% at ignition, and transiently by less than 25\% during startup. It is commonly assumed that the Sudo density limit can be exceeded by 50\% \cite{lyon_systems_2008,peterson_density_2005}, or more if the plasma is very pure \cite{miyazawa_self-sustained_2006}.

For the purposes of 0D scoping, the density and temperature profiles are assumed to follow a parabolic power law, which allows for analytic evaluation of volume-averaged quantities \cite{kovari_process_2014,freidberg_designing_2015}. Impurity and ash dilution is included based on an assumed fraction. Thermal conversion and auxiliary heating efficiencies are assumed based on likely and achieved values. A facility power balance model was developed commensurate with this level of detail. This systems model was used to widely explore the design space of possible scales, magnetic field strengths, and equilibrium types. 

Next, the 1D BP3 code was used to further explore the available operational scenarios \cite{geigerBurning,geiger_prospects_2024}. This includes 1D profiles that are self-consistent with respect to an assumed W7-X-like thermal diffusivity profile shape, and radiation effects. BP3 was used to further refine the power plant operational point.

\begin{figure}[t]
\centering
\includegraphics[width=1.0\linewidth]{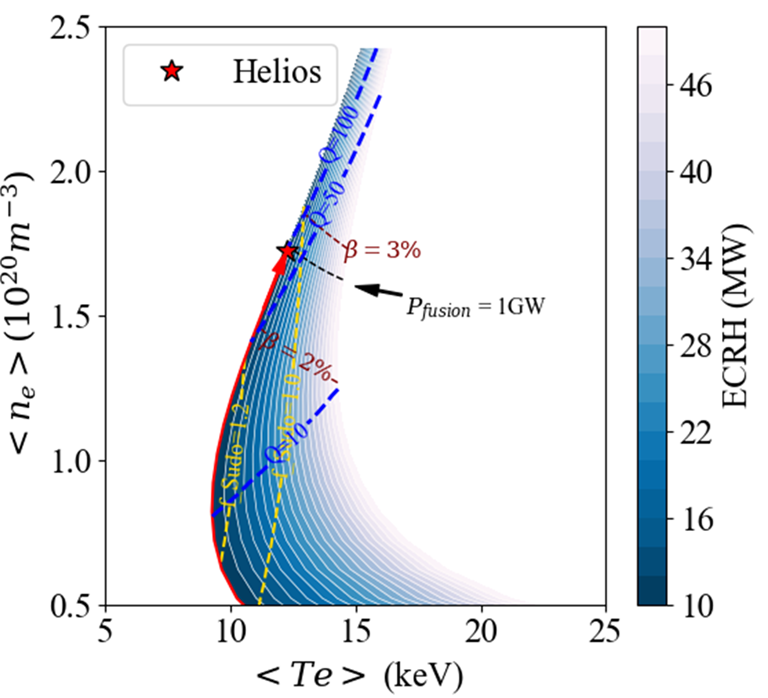}
\caption{POPCON plot of the plasma operational trajectory (red) from zero plasma pressure through an ignited state. 10 MW of ECH power is required to start up. No operational constraints are violated. }\label{fig:popcon}
\end{figure}

BP3 was also used to develop startup scenarios. The POPCON approach is used to visualize the path from zero plasma pressure through an ignited mode. Magnetic control of the plasma during startup is not considered in the POPCON plot but is required. Magnetic control is required to maintain nested flux surfaces, maintain good quasi-symmetry, and ensure that the divertor strike points stay within their dedicated plasma-facing components. Preliminary modeling suggests that the planar coil set described in Section \ref{sec:coils_phys} is sufficient to provide this control. This startup procedure is currently envisioned to occur over a timescale of $\sim2$ hours. 

The POPCON plot is shown in Figure \ref{fig:popcon}. It highlights that only 10 MW of electron cyclotron resonance heating (``ECRH'') power is required to start up Helios. The plasma starts up in a hot and tenuous mode, then densifies slowly over time via additional core and edge fueling. The Sudo density limit is never exceeded by more than 25\%, and in the ignited state is exceeded by less than 10\%. The beta is never higher than 2.7\%, which is enforced as a hard limit for conservatism. Stability at this beta is confirmed in Section \ref{sec:MHD}. Figure \ref{fig:popcon} differs from a typical POPCON plot because only the thermally stable branch is shown. Typically, POPCON plots include an unstable cooler, denser branch, but this branch does not appear in the initial-value BP3 code because it is thermally unstable. 

Heating occurs via ECRH at 170 GHz using ITER-spec gyrotron tubes \cite{ikeda_progress_2021}. The microwaves are launched from the high-field side in the X1 polarization. 10 MW of heating is required during startup operation, then only nominal heating (1 MW) is required once the plasma self-heats in the ignited phase, to expel impurities from the core \cite{ford_turbulence-reduced_2024}. Fueling occurs via deuterium and tritium ice pellet injection and edge gas puffing. Hydrogenic ice pellet injection has occurred in plasma physics experiments, using a single isotope species and for limited-duration discharges \cite{dibon_blower_2015}. Edge gas puffing is required to maintain edge conditions suitable for the divertor. 

For more information on the scoping studies, heating and fueling, and dynamic accessibility of Helios, see the dedicated companion paper in this special issue: \cite{kumar_heating_nodate}.

\subsection{The stellarator equilibrium}
\label{sec:equilibrium}
A two-field-period, quasi-axisymmetric equilibrium was developed to serve as the reference equilibrium for Helios. The preliminary targets for the scale and parameters of the equilibrium were determined using 0D scoping models discussed in Section \ref{sec:scoping}. The DESC stellarator optimization suite was used to represent the equilibrium and optimize it for techno-economic (including plasma physics) figures of merit \cite{dudt_desc_2020}. 

These figures of merit included (in part) quasi-symmetry for neoclassical particle confinement \cite{helander_theory_2014}, the Mercier criterion for ideal MHD stability \cite{mercier_equilibrium_1964}, the ideal ballooning growth rate \cite{Gaur2024}, and consistency with the pressure-driven bootstrap current \cite{landreman_optimization_2022}. A set of encircling coils were co-optimized along with the plasma equilibrium in so-called single-stage optimization \cite{jorge_single-stage_2023}. A shell of scalar current potential \cite{merkel_solution_1987,landreman_improved_2017} was co-optimized, standing in for the shaping coils. In this manner, metrics of coil feasibility were directly targeted in the equilibrium, including proxies for magnetic field strength, stress, and total coil cost. 

\begin{figure}
\centering
\includegraphics[width=1.0\linewidth]{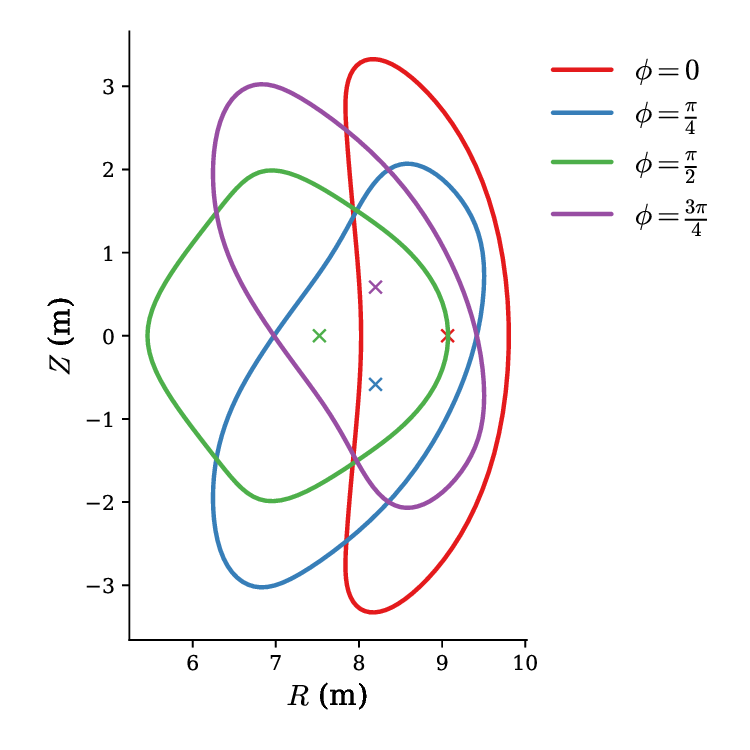}
\caption{Four toroidal cross-sections of the boundary of the Helios plasma equilibrium. The equilibrium was optimized for several techno-economic figures of merit including proxies for coil cost and complexity.}\label{fig:boundary}
\end{figure}

The resultant equilibrium is shown in 3D rendering in Figure \ref{fig:coils_eq}, and four toroidal cross-sections of the boundary can be seen in Figure \ref{fig:boundary}. It has 8 m major radius, an aspect ratio 4.5, and 6.0 T magnetic field strength on-axis. The value of beta is 2.7\%. The maximum value of rotational transform occurs at outer-mid-radius and is $\iota_{max} = 0.46$, of which roughly $1/3$ is due to vacuum field shaping and $2/3$ is due to bootstrap current.

While Figures \ref{fig:coils_eq} and \ref{fig:boundary} depict the fixed-boundary target equilibrium, four near-identical equilibria are actually analyzed in this report. The fixed-boundary equilibrium is used to model MHD behavior and turbulent transport (Sections \ref{sec:MHD} and \ref{sec:turbulence}), and to design electromagnetic coils (Section \ref{sec:coils_phys}). An identical equilibrium with the same beta but hotter and more tenuous plasma was developed using the BP3 code during scenario development (Section \ref{sec:scoping}). A free-boundary equilibrium fit to the reference coil set (Section \ref{sec:coils_phys}) was used for energetic fusion product confinement (Section \ref{sec:alphas}). A second free-boundary equilibrium fit to a re-optimization of the reference coil set was used to study the divertor (Section \ref{sec:divertor_phys}), as the X-point of the reference coil set is 10 cm further from the plasma boundary. The fusion product confinement of the free-boundary of the re-optimized coil set is similar to that of the reference coil set. 

For more information on the design and properties of the stellarator equilibrium, see the dedicated companion paper in this special issue: \cite{dudt_equilibrium_nodate}. 

\subsection{Energetic particle confinement}
\label{sec:alphas}
Confinement of energetic fusion products is not guaranteed in stellarators; rather it must be directly optimized for via one of several proxies \cite{helander_theory_2014,nemov_evaluation_1999,nemov_poloidal_2008}. For Helios, quasi-symmetry was found to be the most effective at producing equilibria which confine energetic particles. 

The ASCOT5 code was used to simulate the behavior of energetic fusion products within the Helios equilibrium, including collisions \cite{varje_high-performance_2019}. A free-boundary equilibrium was used to include the effect of discrete coils. 6.6\% of the fusion product energy is simulated to be lost to the wall. While this is higher than some examples in academic literature \cite{landreman_magnetic_2022}, it is entirely sufficient for plasma self-heating and ignition. The energetic particle power deposition to the walls is highly peaked, with some areas receiving up to 4 MW/m$^2$ of heat flux; this is an area of ongoing optimization. 

Diffusive drift is the dominant loss mechanism. The majority of lost alpha orbits exhibit significant variation in $J_{||}$ associated with diffusive drift \cite{paul_energetic_2022}, Further optimizations will target this loss channel.

For more information on the confinement of the energetic fusion products, see the dedicated companion paper in this special issue: \cite{von_der_linden_alpha_nodate}. 

\subsection{Magnetohydrodynamic stability and evolution}
\label{sec:MHD}
The magnetohydrodynamic (``MHD'') properties of the Helios plasma have been evaluated using the ideal, linear, spectral stability code TERPSICHORE \cite{anderson_terpsichore_1990,anderson_methods_1990} and the resistive, nonlinear, time-domain evolution code M3D-C1 \cite{jardin_multiple_2012,zhou_approach_2021,saxena_bootstrap_2025}. The TERPSICHORE results are generally consistent with stability, though interpretation is required. No large-scale unstable mode is seen in the M3D-C1 simulations. 

The growth rate of the most unstable mode found by TERPSICHORE is positive at the operating beta, with a value of $\gamma/\omega_A = 1.42\%$ where $\omega_A$ is the Alfv\'{e}n frequency. The interpretation of TERPSICHORE results is non-trivial but best practices have been developed by comparing the results to stellarator experiments \cite{turnbull_ideal_2011}. A mode is typically considered serious if its growth rate exceeds 2\% of the Alfv\'{e}n frequency. The Helios equilibrium at the operational beta has a most-unstable growth rate below this value, thus we move on to a higher-fidelity code to further characterize the MHD behavior. 

\begin{figure*}
\centering
\includegraphics[width=0.8\linewidth]{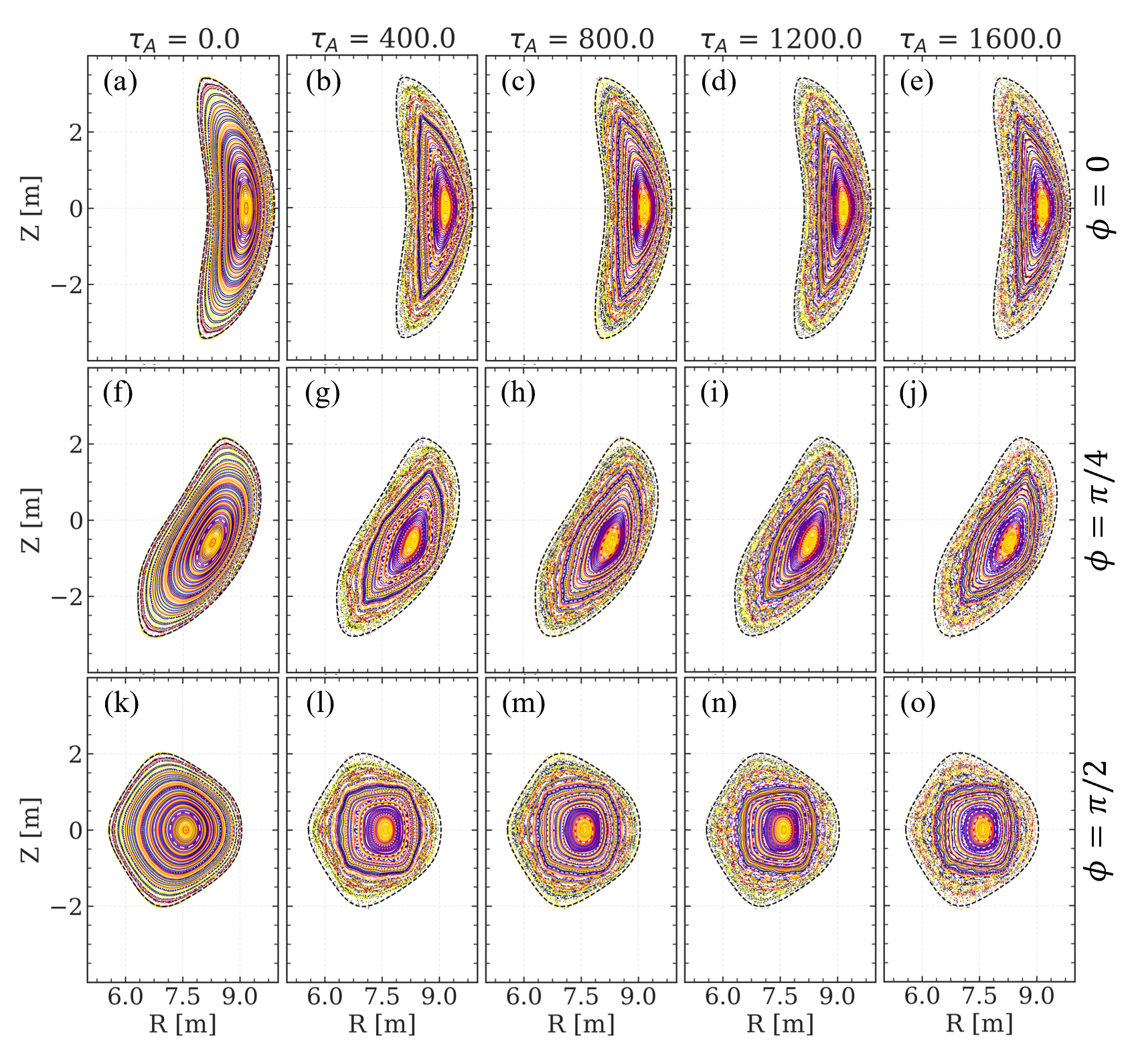}
\caption{M3D-C1 simulation results. Poincar\'e sections of the magnetic field at $\phi=0$ (a-e), $\pi/4$ (f-j), and $\pi /2$ (k-o) shown as a function of time for $0 \leq t \leq 1600\tau_A $ (left to right). Perfect stability would be indicated by nested flux surface of different colors, and little change between times. Small island chains do appear. Stochastic regions occur at the edge but do not result in pressure profile flattening. No large-scale unstable modes are in evidence.}\label{fig:mhd_poincare}
\end{figure*}

M3D-C1 is a resistive, nonlinear, time-domain, MHD evolution code first developed for tokamaks \cite{jardin_multiple_2012} and later extended to stellarators \cite{zhou_approach_2021}. We have recently added a model of stellarator bootstrap current to the code \cite{saxena_bootstrap_2025}. A perfectly conducting wall is placed 10 cm from the plasma boundary. A ratio of $\kappa_\perp / \kappa_\parallel =10^{-5}$ is used, a realistic anisotropy of the thermal diffusivity. Poincar\'e sections of the magnetic field at three toroidal locations and five times are shown in Figure \ref{fig:mhd_poincare}. 

No large-scale instability is seen. The fluid kinetic energy (not plotted) has an initial spike as the idealized DESC equilibrium relaxes, then decreases over time indicating no growing mode. Some stochastization of the magnetic field can be seen at the edge, but the pressure profile (not plotted) remains peaked even at this high thermal diffusivity anisotropy, indicating that the stochastization is not a major contributor to transport. 

For more information on the ideal MHD stability and nonlinear MHD evolution of the plasma within Helios, see the dedicated companion paper in this special issue: \cite{martin_mhd_nodate}. 

\subsubsection{A note on the effects of an abrupt plasma termination in Helios}

While Helios is designed and simulated to be stable, any plasma may suddenly stop. Off-nominal scenarios can not be prevented with absolute certainty; for example, objects can fall through the plasma (``UFOs'', often wall tiles) and end the discharge \cite{gaspar_thermal_2024}. It is important to note that an abrupt termination in Helios would resemble a radiative collapse of a stellarator plasma more than a potentially damaging disruption of a tokamak plasma. Helios is designed not to take damage during a termination, and be easily re-started. 

Unlike in a stellarator, the stored magnetic energy in a tokamak plasma due to the plasma current is larger than the thermal energy within the plasma. In ITER, there is approximately 1 GJ of plasma magnetic energy and 350 MJ of plasma thermal energy \cite{hollmann_status_2014}. This GJ of magnetic energy concentrates nonuniformly as eddy currents in conductive structures and causes large Lorentz forces. ITER is designed to survive only 15 un-mitigated disruptions due to this challenging dynamical system. 

In Helios, the scenario is very different. The Helios equilibrium carries plasma current due to neoclassical effects (bootstrap current), but it does not approach that of a similarly scaled tokamak. The much smaller JET tokamak experiment routinely carried more plasma current than the power-plant-scale Helios. The plasma magnetic energy from the bootstrap current is only approximately 100 MJ, 10\% that of ITER. The plasma thermal energy is double this, making the dynamics more stellarator-like than tokamak-like. Because of the existence of vacuum rotational transform, it is likely that the plasma current is magnetically confined even in the case of an abrupt termination. While modeling abrupt termination is left to future work, this difference in scale and kind give confidence that it would not damage the stellarator system. 

For more discussion, see the MHD companion paper in this special issue: \cite{martin_mhd_nodate}. 

\subsection{Turbulence, transport, and profile prediction}
\label{sec:turbulence}
For the purposes of scoping and scenario development (see Section \ref{sec:scoping}), transport in Helios was assumed to follow the ISS04 scaling \cite{yamada_characterization_2005}, with a confinement enhancement multiplier of $H_{ISS04} = 1.4$. This value has been achieved in the W7-X stellarator\cite{beurskens_ion_2021,bozhenkov_high-performance_2020}. However this level of turbulent transport should also be justified via high-fidelity, first-principles simulation. 

The GENE code was used to simulate electrostatic gyrokinetic evolution within a flux tube \cite{jenko_electron_2000}. This analysis produces the local plasma heat flux. This calculation was coupled to the Trinity 3D (``T3D'') code \cite{qian_stellarator_2022}, an extension of the Trinity code \cite{barnes_trinity_2009,barnes_direct_2010} to the stellarator geometry. This code evolves the temperature profile across a larger temporal and spatial domain than the gyrokinetic simulation, efficiently reaching a self-consistent, steady-state profile. Additionally, neoclassical transport, electron-ion collisional equilibration, auxiliary heating, fusion product heating, and radiation are considered by T3D. The density profile is fixed, as is the temperature boundary condition at the outer edge, $\rho=0.85$. 

This exploration resulted in a scenario which was developed with an identical fusion power and slightly lower confinement multiplication factor of $H_{ISS04} = 1.33$. The density and auxiliary power are commensurately increased to compensate. The ion temperature is unchanged from the reference case, but the electron temperature is increased by 5 keV in the core due to its preferential heating via electron cyclotron resonance and fusion product heating. The profile shapes are otherwise broadly similar. 

The shape of the Helios equilibrium rotational transform $\iota(\rho)$ has a wide region of what tokamaks would call reversed or negative magnetic shear $-\rho \partial_\rho \iota/\iota > 0$, which has been shown to strongly suppress turbulence in experiments \cite{garofalo_access_2006,kessel_improved_1994,taylor_optimized_1994}. This raises the intriguing possibility that Helios would exhibit a much higher $H_{ISS04}$, requiring less heating and/or permitting a smaller device to produce net electric power. 

For more information on transport, turbulence, and profile prediction in Helios, see the dedicated companion paper in this special issue: \cite{martin_profile_nodate}.

\subsection{Electromagnetic coil physics design}
\label{sec:coils_phys}

\begin{figure}
\centering
\includegraphics[width=1.0\linewidth]{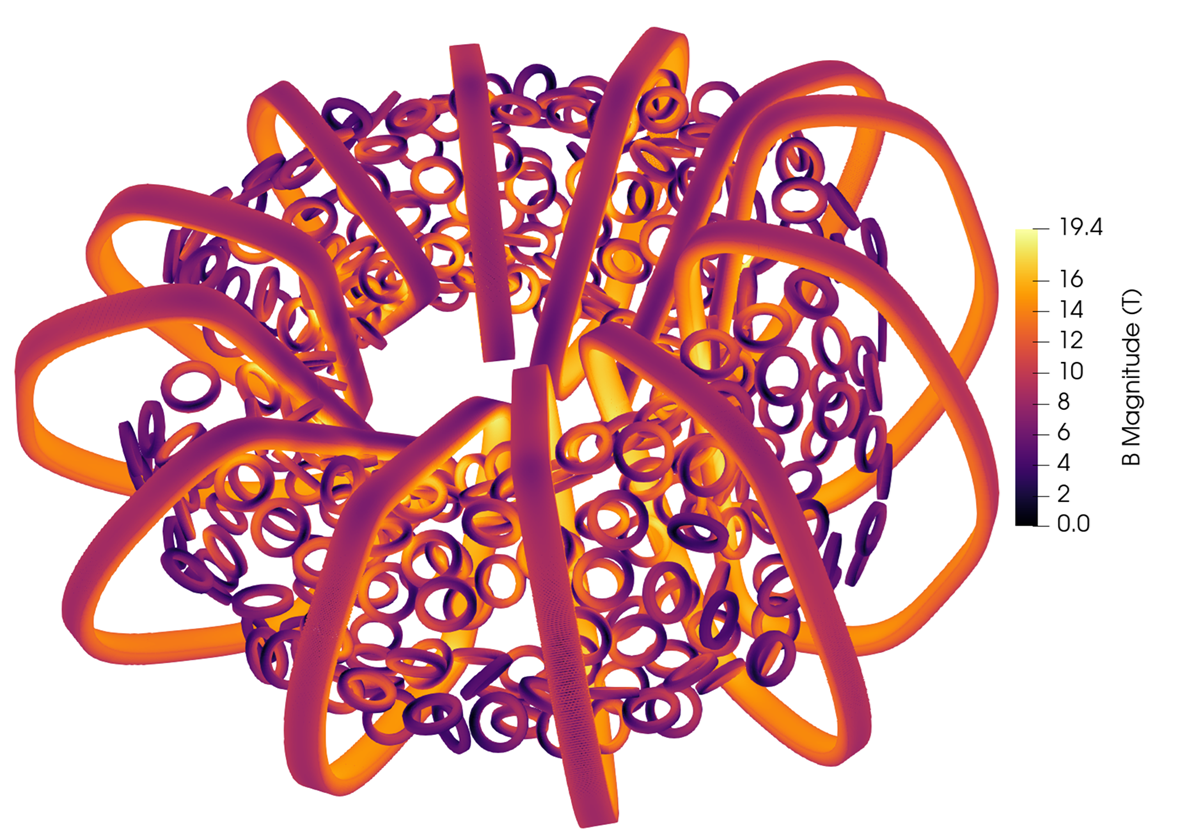}
\caption{Rendering of the planar coil set of Helios, shaded by the magnetic field strength $|B|$ on the winding pack. No superconductor is subjected to a magnetic field stronger than 20 T.}\label{fig:coils_B}
\end{figure}

Helios uses a novel planar coil architecture consisting of 12 planar, plasma-encircling coils similar to the toroidal field coils of a tokamak, and 324 smaller, planar, field-shaping coils that surround the plasma boundary. All shaping coils are circular and have the same radius, minimizing the number of unique parts. Thea Energy has previously published on the advantages of this coil set \cite{gates_planar_2024,gates_stellarator_2025,swanson_scoping_2025,kruger_coil_2025,wu_planar_2025}. All coils are planar and convex, and can therefore be wound in tension. Hundreds of individually controllable coils permit an unprecedented degree of magnetic field control and configurability. This configurability permits much looser manufacturing and assembly tolerances, as errors can be tuned out during operation by controlling the coil currents. The large gaps between the encircling coils permit entire toroidal sectors of the radial build to be removed for maintenance. 

As discussed in Section \ref{sec:equilibrium}, the process of single-stage stellarator equilibrium optimization also produces an initial guess for a set of encircling coils, and a shell of scalar current potential that approximates a set of shaping coils. These are used as the initial guesses in a second stage of coil optimization, which has been described previously \cite{kruger_coil_2025}. The coil winding packs are shown, along with their on-tape magnetic field strength, in Figure \ref{fig:coils_B}. 

In the optimization, the usual quadratic flux penalty ($\int dA (\vec{B}\cdot\hat{n})^2$) is used to ensure that the equilibrium is accurately reconstructed. The average normal field error on the equilibrium boundary is 0.21\%. The good reconstruction of a quasisymmetric magnetic field can be seen by the $|B|$ contours on the target (fixed-boundary) equilibrium boundary in Figure \ref{fig:coils_eq}. The equilibrium is reconstructed sufficiently well that energetic particle confinement is good in a free-boundary equilibrium fit to this coil set; see Section \ref{sec:alphas}. 

By using the recent approximate formulation of Landreman \textit{et al.} \cite{landreman_efficient_2025}, magnetic field on-coil can be efficiently calculated, and quantities such as maximum field on-coil and total HTS tape length may be optimized directly. This field on-coil is shown in Figure \ref{fig:coils_B}; the maximum field on-coil is 20 T, limiting quench energy and stress. Large-bore, HTS coils have been constructed and operated at 20 T, indicating achievable engineering feasibility \cite{hartwig_sparc_2023}. 

A minimum distance of 1.2 m between the plasma and coils is enforced. This leaves space for adequate breeding blanket and neutron shielding; see Section \ref{sec:blanket}. A novel technique is used to enforce the existence and location of a tokamak-like X-point divertor; see Section \ref{sec:divertor_phys}. Free-boundary plasma equilibria are used for the plasma physics analyses, so that the effect of finite coils is included. 

The engineering of these coils is described in more detail in Section \ref{sec:coils_eng}. For more information on the planar coil architecture and the physics design of the Helios electromagnetic coil set, see the dedicated companion paper in this special issue: \cite{kruger_planar_nodate}. 

\subsection{Divertor physics}
\label{sec:divertor_phys}

Helios features a novel toroidally continuous non-resonant X-point divertor, like that of a tokamak. For a tutorial approach to the tokamak poloidal divertor, see Chapter 5 of Stangeby's textbook \cite{stangeby_plasma_2000}. This type of divertor has been theorized in stellarators, and recently discovered in a database of QA vacuum solutions \cite{davies_topology_2025}. This is the first time to our knowledge that a stellarator equilibrium has explicitly designed to include such a divertor, and the first time that a power-plant-relevant equilibrium with finite beta and bootstrap current has been paired with such a divertor. It is the most detailed exploration of the practicality of such a divertor.

This novel divertor type has the potential to resolve an outstanding problem in stellarator design. Several divertor types have been proposed, well summarized in Chapter 2.3 of the Stellcon report \cite{gates_stellarator_2018}, but none have a clear scaling to a power plant system. The experimentally verified frontrunner is arguably the island divertor, as implemented on W7-X \cite{sunn_pedersen_first_2019}. However, the island divertor is acknowledged by its proponents not to scale directly to a power plant as its neutral particle compression is insufficient to enable a practical vacuum pumping scheme \cite{lion_stellaris_2025,hegna_infinity_2025}.

In tokamaks, the X-point divertor has been modeled to compress plasma density at the target an order of magnitude more effectively than the island divertor \cite{feng_comparison_2011}. This is counterintuitive from the simple two-point model perspective \cite{stangeby_plasma_2000}; the increased connection length of the island divertor would appear to be superior. However the inclusion of perpendicular transport into an extended two-point model \cite{feng_comparison_2011} reveals that the very low pitch angle that accompanies a high connection length causes saturation of the compression effect. 

\begin{figure}[t]
\centering
\includegraphics[width=1.0\linewidth]{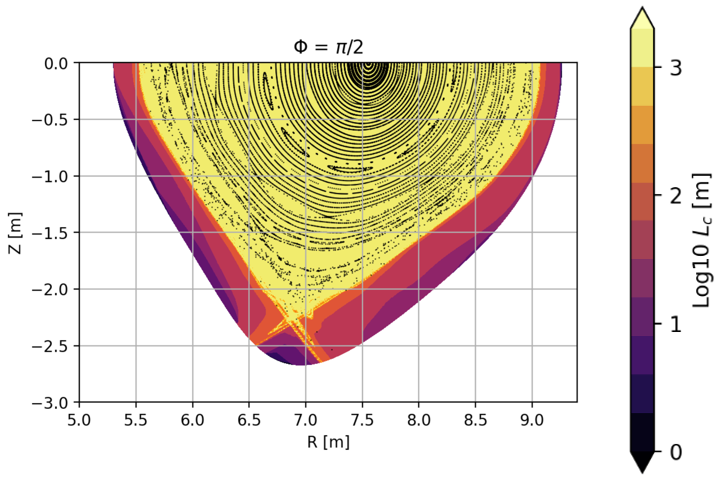}
\caption{Poincar\'e section (black) and connection length (shaded) plots for a toroidal cross-section of the equilibrium. Features isomorphic to a tokamak poloidal divertor topology are present, including divertor legs and a private flux region. The divertor stays on the bottom of the plasma for the entire toroidal domain.}\label{fig:x_point}
\end{figure}

Additionally, the tokamak-like divertor is more easily baffled, enabling greater compression of the neutral particles born when the plasma contacts the divertor target. From a practical standpoint, the tokamak-like X-point divertor is simpler in that it lies on the bottom of the plasma and does not multiply link it as do competing divertor designs. 

The FLARE code \cite{frerichs_flare_2024} has been used to explore the topology and behavior of this novel divertor in Helios. The equilibrium used is a free-boundary fit to the planar coil set, and self-consistent plasma current is included. A toroidal cross section of this of the wall-to-wall connection length can be seen in Figure \ref{fig:x_point} and compared to the familiar tokamak poloidal divertor. This connection length was used to place divertor target elements and neutral gas baffles. 

A simple field line diffusion model was used to estimate heat flux on the targets, again using FLARE. The result shows that additional consideration is required to keep the heat flux below an assumed limit of 10 MW/m$^2$; some combination of radiative impurity seeding in the edge, or detachment, or enhanced core radiation, or finely contoured targets to maximize the strike point uniformity.

The engineering design of this divertor, including plasma-facing components, is described in Section \ref{sec:divertor_eng}. For more information on the magnetic topology and early plasma physics modeling of this tokamak-like X-point divertor, see the dedicated companion paper in this special issue: \cite{flom_design_nodate}.

\section{Engineering design of systems}
\label{sec:eng}

In this section, the engineering design of the systems that generate the plasma and receive the fusion power are discussed in more detail. A strong driver of the design is a facility that is buildable, and maintainable under unforeseen circumstances. Each subsection discusses a system. The reader is referred to the other papers in this special issue for more detail on each subject. 

In Section \ref{sec:coils_eng}, the coil system, including quench and structure, is described. In Section \ref{sec:divertor_eng}, the divertor, first wall, and vacuum systems are described. In Section \ref{sec:blanket}, neutronic analysis of the blanket and shield is presented. In Section \ref{sec:cycles}, the thermal cycle, fuel cycle, and the power flow diagram for the facility are shown. In Section \ref{sec:maintenance}, the sector maintenance scheme, cryostat, and cryogenic system are discussed. Section \ref{sec:electrical} outlines the electrical system and power supplies. Section \ref{sec:control} discusses the instrumentation and control of the plasma and facility. 

\subsection{Electromagnetic coil engineering}
\label{sec:coils_eng}
As described in Section \ref{sec:coils_phys}, Helios uses a set of planar, plasma-encircling coils and planar, field-shaping coils. 

\begin{figure}[t]
\centering
\includegraphics[width=1.0\linewidth]{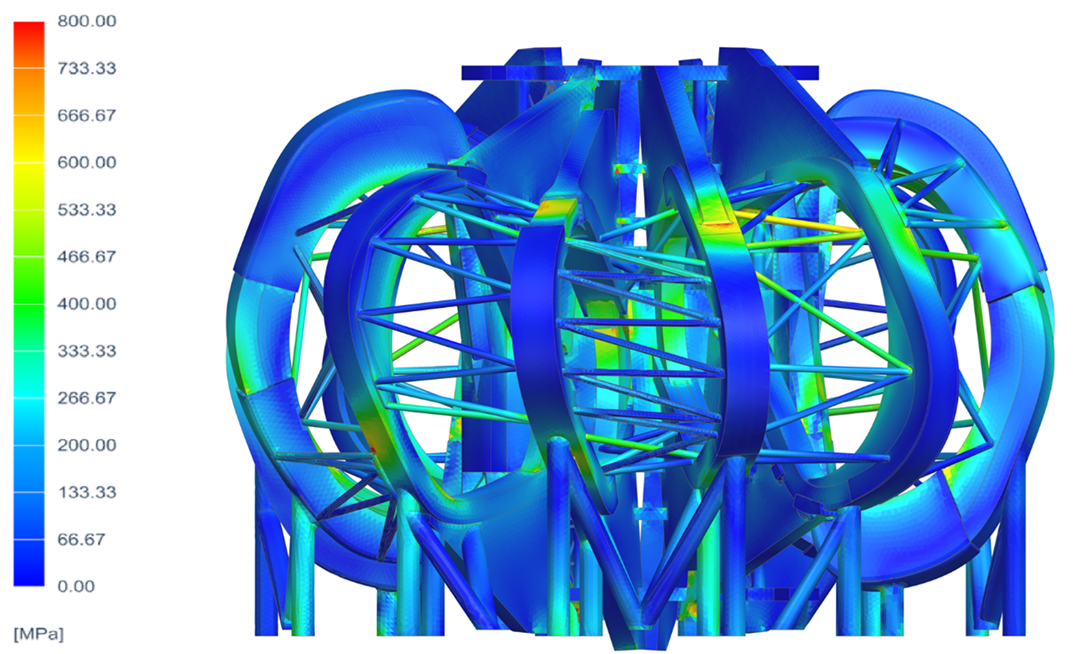}
\caption{Finite element analysis of the von Mises stress distribution within the encircling coil structure. Red: 800 MPa. The coil cases, central support structure, and inter-coil trusses are visible and below acceptable stress limits. The trusses can be removed to access components for maintenance.}\label{fig:coil_stress}
\end{figure}

The 12 encircling coils are composed of pancakes of HTS cable, operated at 20 K. They are similar to tokamak toroidal field coils. Their turns are insulated, and superconductor quench mitigation is active and external via a dump resistor. The structural solution within encircling coils is based on a stainless steel case, providing the stiffness to maintain the on-coil strain at less than 0.4\%. The structural solution between encircling coils includes a central support structure and inter-coil truss structures. The central support structure is analogous to a tokamak bucking cylinder \cite{bromberg_magnet_1990}. The structural solution is compatible with the sector-based maintenance scheme described in \ref{sec:maintenance}. No part of the structure exceeds 800 MPa, compatible with widely available stainless steel alloys. With additional design, a maximum stress not exceeding 600 MPa appears achievable. These reasonable structural requirements are a benefit both of the less-shaped QA equilibrium choice and of the decision not to exceed 20 T on-coil. This structure is shown in Figure \ref{fig:coil_stress}.

The 324 shaping coils are composed of pancakes of partially insulated HTS tape stacks, operated at 20 K. The large number of coils necessitates a high-inductance, low-operating-current approach to minimize the heat leak due to charging cables. The shaping coils all have the same inner and outer diameters, enabling only one size of pancake to be manufactured. These pancakes can be stacked in different numbers to create shaping coils with different magnetic strengths. The coils are self-protecting with respect to quench by virtue of the partially insulated approach; a multi-physics COMSOL analysis simulates this process. Shaping coils are ganged together into field-shaping units (``FSUs'') which can be removed from between the encircling coils. FSUs route services such as cryogens, power, and data to shaping coils. 

For more information on the engineering of the Helios planar coil set, see the dedicated companion papers in this special issue: Shaping coils \cite{olatunji_design_nodate-1} and encircling coils \cite{olatunji_design_nodate}.

\subsection{Divertor engineering and the first wall}
\label{sec:divertor_eng}

\begin{figure}
\centering
\includegraphics[width=1.0\linewidth]{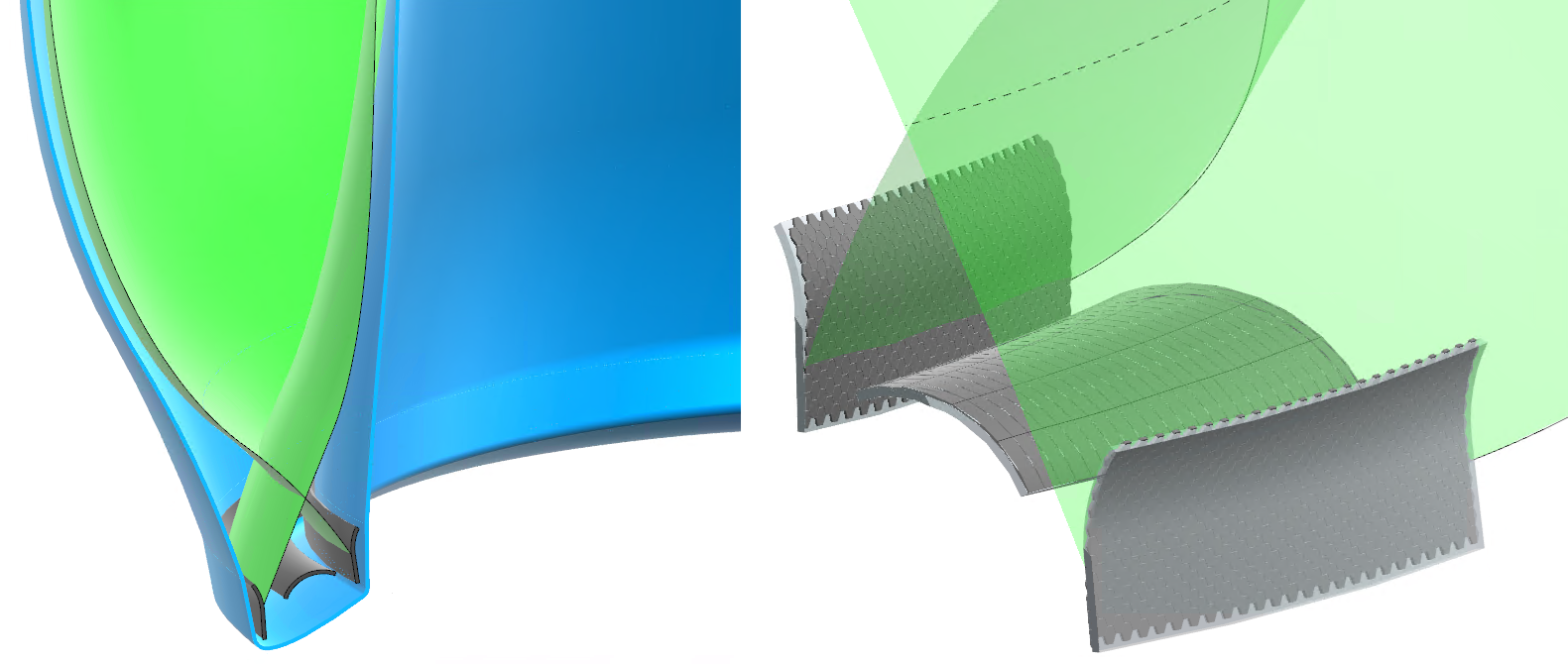}
\caption{Rendering of the divertor system. Left: Postion of the plasma (green) within the first wall (blue), with the targets and dome (grey). Right: Detail of the targets and dome. Pump ducts are not pictured.}\label{fig:divertor_system}
\end{figure}

As described in Section \ref{sec:divertor_phys}, Helios has a novel toroidally continuous non-resonant X-point divertor like that of a tokamak, a first for the design of an optimized stellarator. The divertor system consists of high-heat-flux plasma-facing target plates and an ITER-like dome to improve neutral compression lying below the plasma, as is common in tokamak configurations \cite{kukushkin_effect_2007,lackner_recent_1994,galassi_numerical_2020,reimerdes_initial_2021}. The targets are composed of 51,000 tessellated hexagonal tiles of tungsten 2.5 cm in width, cooled with a closed system of helium impingement jets. Their location within the system can be seen in Figure \ref{fig:divertor_system}. An assumed 10 MW/m$^2$ of heat flux is incident on the targets, using Siemens Simcenter Amesim for thermofluid modeling to design tiles that remain within acceptable temperature limits. The hexagonal tiles are mounted to a vanadium alloy support structure. The helium coolant enters from the blanket, where it has been pre-heated to above the ductile-to-brittle transition temperature of tungsten.

The vacuum pumping system uses turbomolecular pumps rather than cryosorption pumps due to their efficiency at pumping helium and steady-state operation without regeneration. The pumps themselves must be situated far enough from the stellarator that they can be magnetically shielded \cite{de_angeli_design_2007,wolf_investigation_2011}; this necessitates longer pump ducts than a cryopump based design. The high neutral compression of the tokamak-like X-point divertor enables this design. The dimensions and performance of the pump ducts were designed using lumped-element conductances of choked molecular, transitional, and fluid flow \cite{livesey_flow_1998}.

The first wall is 2 cm thick with integrated helium cooling channels. It is composed of a vanadium alloy layer with a thin layer of tungsten armor. Specifically the alloy considered is V-4Cr-4Ti (``V44''), though there are significant uncertainties with respect to the suitability of materials for fusion environments \cite{el-guebaly_nuclear_2006}. V44 is used in this design exercise over EUROFER97 or other RAFM steels because of its higher operating temperature, no ferromagnetism, high neutron damage tolerance, and low activation properties \cite{el-guebaly_nuclear_2006,smith_vanadium-base_2000,sparks_mechanical_2022,chen_mechanical_2006,li_mechanical_2011,zinkle_thermophysical_1998}. The chief advantage of using V44 in the first wall is its resistance to high-energy neutron damage, permitting the first wall a lifetime of 15 full-power years (compared to less than half that for reduced-activation steels). This is discussed in Section \ref{sec:blanket}. The superior high-temperature creep strength of V44 allows for an additional factor of safety over RAFM steel. Considerations potentially contraindicating V44 include its high affinity for hydrogenic species, swelling under irradiation, and immature supply chain. 

For more information on the design of the Helios divertor systems and the first wall, see the dedicated companion paper in this special issue: \cite{kalb_preliminary_nodate}.

\subsection{Neutronics, blanket, shield, and bioshield}
\label{sec:blanket}

The reference design for the tritium breeding blanket uses a lead-lithium breeding medium. This is one of the more common types of blanket architectures. It is the choice of the ARIES-CS, EU-DEMO, and recent Stellaris studies, among others \cite{el-guebaly_designing_2008,aubert_development_2014,lion_stellaris_2025}. In it, Li-6 nuclei within a flowing molten alloy of lead and lithium, usually at the eutectic mixture of 17\% lithium by atom (``Pb-17Li''), absorb neutrons and produce tritium via the $^6\textrm{Li}+\textrm{n}\rightarrow {^3\textrm{H}}+{^4\textrm{He}}$ ``breeding'' reaction. Typically, the lithium in the mixture is required to be isotopically enriched with lithium-6. Typically, an additional coolant fluid other than the liquid metal is required to remove heat at a sufficient rate. 

The Helios lead-lithium blanket design is enriched with 65\% lithium-6 isotope. It completely surrounds the plasma and first wall in a uniform 50-cm-thick layer. EUROFER97 \cite{gaganidze_mechanical_2011} is used as the structural material and cooled with helium gas. The lead-lithium flows at a rate of 6.6 cm/s; it is then run through a heat exchanger and the tritium extraction system and returned to the blanket. Silicon Carbide (``SiC'') inserts form an inert and non-conductive break between the flowing lead-lithium and the EUROFER97 structure, reducing the MHD pressure drop and pumping power to a reasonable level. The tritium breeding blanket is split into sectors that fit between the encircling coils and allow for sector-based maintenance, as discussed in Section \ref{sec:maintenance}.

\begin{figure}[t]
\centering
\includegraphics[width=1.0\linewidth]{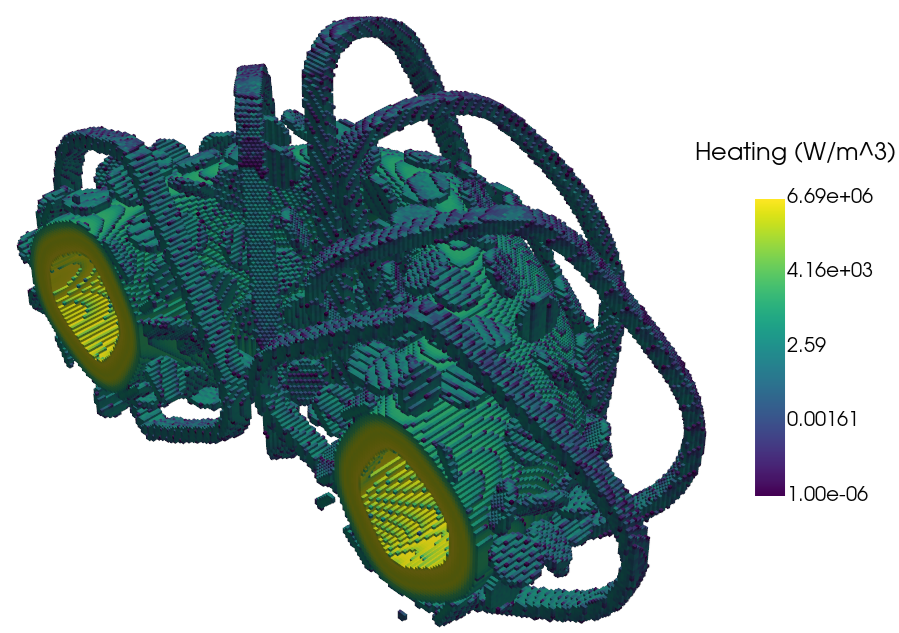}
\caption{Cutaway of volume-resolved nuclear heating density as simulated by OpenMC. The heating is highest at the first wall and decays through the breeding zone and shield.}\label{fig:blanket3D}
\end{figure}

The OpenMC Monte Carlo neutron and photon transport simulation code was used to calculate the neutronic properties of the blanket \cite{romano_openmc_2015}; A novel implementation of the cell-under-voxel algorithm was written and contributed to OpenMC in order to perform volume-resolved simulations \cite{romano_computing_2025}. OpenMC was used to predict the neutron flux, damage rate, tritium yield, material activation, nuclear heating, photon activity, and effective dose in the first wall, blanket, shield, coils, and in the stellarator hall. A cutaway rendering of the nuclear heating in the structure is given in Figure \ref{fig:blanket3D}.

The idealized (homogenized, no penetrations) tritium breeding ratio (``TBR'') is 1.3, a good target for this level of fidelity to allow for non-ideal reductions \cite{el-guebaly_designing_2008}. Fuel cycle modeling indicates that a final TBR of 1.1 is sufficient for tritium self-sufficiency of the facility. 

As discussed in Section \ref{sec:coils_phys}, a minimum distance of 1.2 m between the plasma and coils was rigorously enforced. This distance permits a thick neutron shield between the breeder zone and the coils to protect them from the damaging high-energy neutron flux. A multi-layer shield was designed to absorb and moderate neutrons of successively lower energies. Closest to the breeding zone is a tungsten carbide (``WC'') shield for high-energy neutrons, followed by successive layers of boron carbide (``B$_4$C''), 316L stainless steel constituting the vacuum vessel, a layer cooled by borated water, and finally borated high-density polyethylene (``HDPE'') for the lowest-energy neutrons. This thick, multi-layer shield strongly attenuates the flux of high-energy neutrons and allows the coils to last more than 40 years, making them lifetime components. The fast-fluence limit of the coils was estimated using measurements of ReBCO HTS exposed to neutrons from a fission core \cite{prokopec_suitability_2014,fischer_effect_2018}.

An approximately 2.0 m thick concrete bioshield ensures the effective dose outside the stellarator hall is below permitted limits for public exposure by the Nuclear Regulatory Commission.

For more information on the design of the Helios blanket, shield, and bioshield, see the dedicated companion paper in this special issue: \cite{pasmann_2025_preliminary}.

\subsection{Thermal cycle, power flows, and fuel cycle}
\label{sec:cycles}

A Rankine cycle for power generation has been designed for Helios. 1.1 GW of thermal power is generated in Helios, both from the fusion reaction itself (958 MW) and from the exothermic $^6\textrm{Li}+\textrm{n}\rightarrow {^3\textrm{H}}+{^4\textrm{He}}$ breeding reaction within the blanket. Temperatures inside the stellarator are carefully controlled to maximize performance and remain within material capabilities. Power from the blanket, first wall, and divertor is transported via lead-lithium and helium to intermediate heat exchangers. These heat exchangers are part of a closed cycle loop that transfers heat to boil water into steam while simultaneously cooling the blanket and helium fluids to their inlet temperatures. The steam is superheated to 635 $^\circ$C, and enters a series of three steam turbines that generate electric power. A simulation of the Rankine cycle design outlines intermediate flow splits and recombination to increase system efficiency and maximize cycle power generation. The total generated power is 460 MW$_e$, with 22 MW$_e$ required to pump the various coolants, for a gross electric power of 438 MW$_e$ from the thermodynamic cycle. The combined efficiency of the Rankine cycle is approximately 40.2\%. If instead the efficiency is computed using the total generated power, as in some fusion power plant design studies, the efficiency is approximately 42.2\%. 

\begin{figure*}
\centering
\includegraphics[height=0.53\linewidth, angle=90]{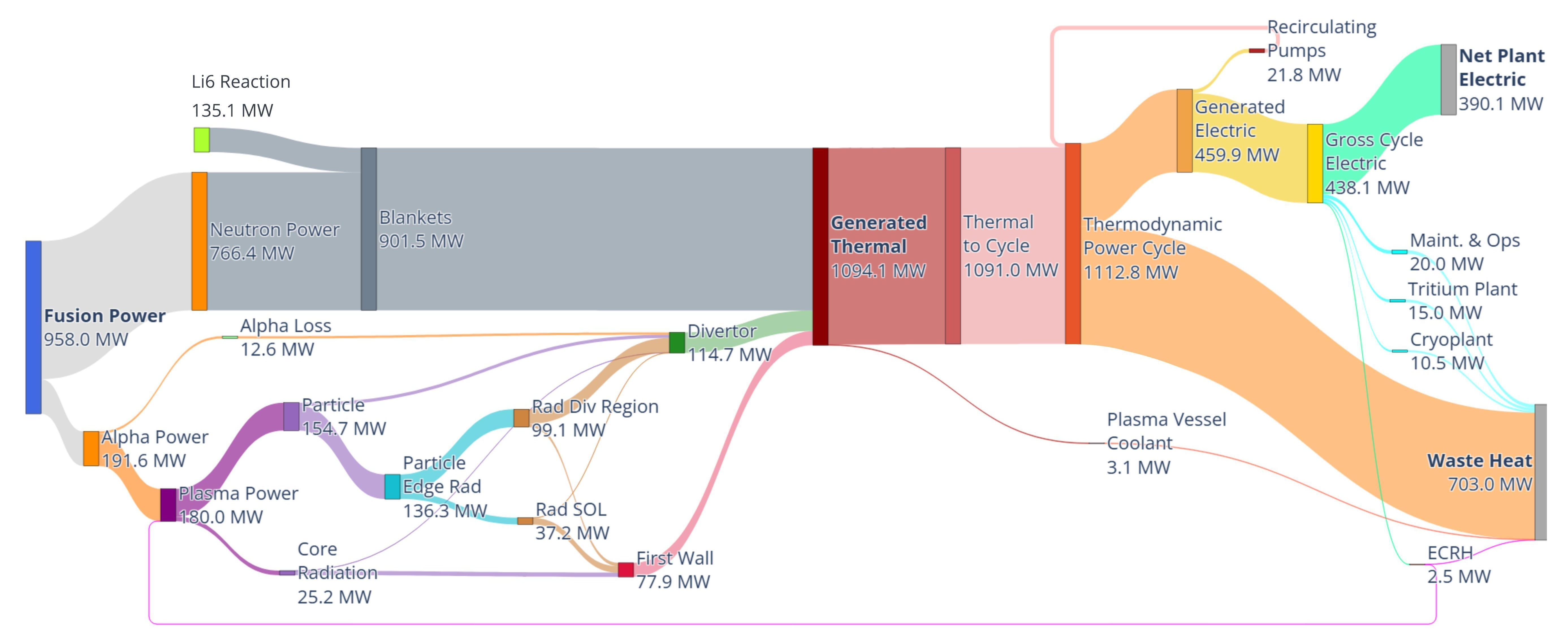}
\caption{Sankey diagram of the facility power flows during ignited operation.}\label{fig:sankey}
\end{figure*}

The global facility power flow is represented by a Sankey diagram in Figure \ref{fig:sankey}. During steady-state, ignited, power-producing operation, 958 MW of power is released within the plasma via deuterium-tritium (``D-T'') fusion. Of this energy, 80\% is released as high-energy neutrons that heat the blanket. There, the exothermic tritium breeding reaction of lithium-6 absorption (less endothermic reactions) produces an additional 135 MW; lithium-6 is also a power-producing fuel.

Of the 192 MW of fusion power that is released as alpha particles within the plasma, 13 MW is lost to the wall without thermalizing. The remainder heats the plasma, and makes its way eventually to the wall as light (electromagnetic radiation from core, edge, and divertor) or particle kinetic energy. In total, 1,094 MW of thermal energy is generated in the plasma and blankets.

Of this thermal power, 460 MW is converted into electrical power via the thermal conversion cycle, of which 22 MW are used to pump coolants, for a gross electric power of 438 MW. 681 MW is lost as waste heat. Of the gross electric power, 48 MW are required to maintain the facility in power producing state, with similar amounts being required to separate tritium fuel, keep the cryogenic magnets cold, and other uses. 2.5 MW is budgeted for a nominal amount of ECRH (1 MW) for plasma control and to expel impurities from the core. This leaves approximately 390 MW of net electric power available to deliver to the grid.

The fuel cycle returns un-fused tritium to storage and extracts tritium bred in the blanket, thence to re-inject into the fusing plasma. We have performed preconceptual design and modeling of a tritium fuel cycle. TMAP8, A network-based model considering flows between reservoirs and residence times of those reservoirs was used \cite{simon_moose-based_2025}. This model, and the fuel cycle analyzed, is typical of ITER, EU-DEMO, and published studies from private companies \cite{clark_breeder_2025,coleman_demo_2019,meschini_modeling_2023,abdou_physics_2020}. The tritium residence times and loss fractions are estimates based on ITER and typical assumptions from literature for each component, and are likely to be of the correct order of magnitude for present technology \cite{abdou_physics_2020}. 

A sensitivity analysis of plant parameters shows that the main drivers of inventory and required TBR include availability factor and doubling time as tradeoff parameters, and tritium burn fraction. fusion power, and reserve time as optimal parameters. Direct internal recycling fraction, breeding zone residence time, and tritium extraction efficiency have minimal impact on system dynamics, but are shown to affect system complexity, blanket inventory, and extraction system inventory respectively. <1 kg of startup tritium and TBR of <1.15 appear satisfactory for tritium self-sufficiency. Technology advances may reduce requirements further. This gives confidence that tritium self-sufficiency is achievable for the Helios FPP with a lead-lithium breeding blanket.

For more information on the thermal cycle, power flows, and fuel cycle of Helios, see the dedicated companion paper in this special issue: \cite{tang_facility_nodate}.

\subsection{Cryostat, maintenance, and cryogenic system}
\label{sec:maintenance}

\begin{figure*}
\centering
\includegraphics[width=0.8\linewidth]{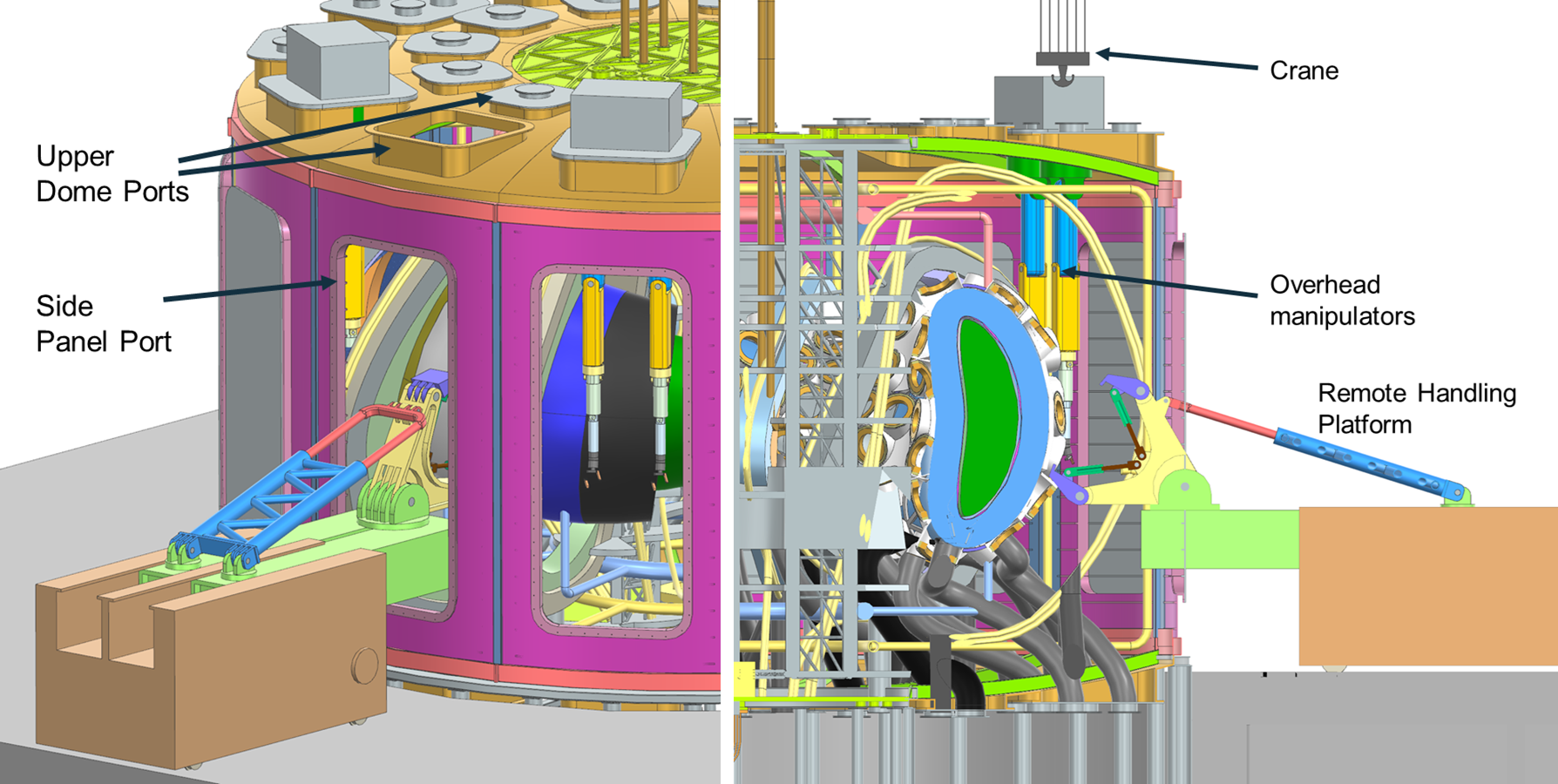}
\caption{Two views of the ground based remote handling platform grappling a sector of the plasma vessel for removal and maintenance. The remote handling platform is extended through a side panel port, with overhead manipulators inserted through adjacent upper dome ports.}\label{fig:maintenance}
\end{figure*}

The stellarator is encased in a stainless-steel cryostat, maintaining a vacuum to limit heat leak into the cryogenic components. The cryostat has large removable radial ports through which components are removed and maintained, in accordance with the sector maintenance scheme. The heat leak and thermal load is about 40 kW into the 20 K cold mass, and about 750 kW into the 77 K thermal shields. Cooling is distributed via helium and nitrogen respectively. There are nominal loads on a 4.5 K system if low-temperature superconducting magnets are used for the gyrotron tubes; these are handled locally. A large expander cycle cryogenic plant to cool these loads would require 10 MW of electrical power, taking a plausible 25\% of the Carnot efficiency.

Helios has a practical maintenance strategy where entire sectors of the radial build are removed, including first wall, blanket, shield, plasma vessel, and shaping coils, from between the encircling coils. This is depicted in Figure \ref{fig:maintenance}. This permits rapid removal and replacement of high-wear layers allowing for a high capacity factor of $\sim$88\%. The encircling coils remain integrated into the stellarator during maintenance. The projected maintenance cadence is one 84-day planned outage every two years.

This is in contrast to alternative stellarator maintenance schemes proposed. ARIES-CS considered a port-based scheme, in which 222 parts are extracted through three small ports \cite{waganer_aries-cs_2008}. This scheme was required because only small spaces exist between the modular coils of the ARIES-CS design. Other studies have considered a field-period-based approach, sometimes called sector-splitting \cite{wang_maintenance_2005,lion_stellaris_2025,hegna_infinity_2025}. In this approach, entire field periods, including the coils, are removed from the stellarator and maintained. For the ARIES-CS design, these field periods were 4,000 tons. This approach was not chosen because the practical challenges of removing and reintegrating entire thirds or quarters of the stellarator appear significant. 

The toroidal sectors that are removed from between the encircling coils represent a balance between the port-based and field-period-based approach. The sectors are not so small that hundreds must be removed serially, but not so large that they weigh thousands of tons and are impractical to remove and reintegrate. The encircling coils remain integrated during maintenance.

The sector-based approach used for Helios is inspired by the same concept in tokamaks, in which entire toroidal sectors of the radial build (first wall, divertor, blanket, shield) are removed from between toroidal field coils \cite{waganer_aries-at_2006}. This sector-based approach has been proposed for stellarators before now, but it required the plasma equilibrium to support modular coils whose outer legs are straight \cite{brown_engineering_2015,gates_recent_2017}. The addition of shaping coils, which can be removed and replaced from between the encircling coils, eliminates this constraint. 

For more information on the maintenance scheme, cryostat, and cryogenic systems of Helios, see the dedicated companion paper in this special issue: \cite{wasserman_helios_nodate}.

\subsection{Electrical systems and power supplies}
\label{sec:electrical}

The electrical systems of Helios supply internal plant loads and provide the grid interface. Steady-state operation requires $\sim$70 MW of auxiliary power, mainly for thermal-hydraulics, tritium processing, cryogenics, and maintenance. This demand is distributed via a 34.5 kV medium-voltage (``MV'') backbone fed by a station-service transformer from the 345 kV grid. Six grouped subsystems draw from this backbone: Primary Heat Transfer System (``PHTS''), ECRH and Magnet power supply units (``PSUs''), Cryogenics, Controls \& Computing, Utilities \& Facilities, and Tritium Systems. To stabilize transients, the MV system includes a $\pm$200–300 Mvar Static Synchronous Compensator “STATCOM” or Static Var Compensator (``SVC''), harmonic filters, and a Battery Energy Storage System (``BESS''). At the high-voltage yard, twin 300 MVA transformers connect the Rankine-cycle turbine-generator to the grid, while the station-service transformer supports auxiliaries during start-up. Power-up sequences energize controls, cryogenics, fueling, and ECRH with ramp-rate limits, after which the plant delivers up to 390 MW$_e$ net electric power to the grid. This integrated architecture ensures reliable auxiliary supply, smooth import-to-export transitions, and high-quality grid delivery.

The power supplies of Helios sustain the stellarator's magnetic configuration and heating through three families: encircling coil, shaping coil and ECRH. Encircling coil supplies drive the 12 superconducting coils with modular converters up to 50 kA, offering four-quadrant control, rapid constant power/current charging, and regenerative discharge with quench protection. Shaping coil supplies regulate the 324 planar coils via modular DC/DC converters with hot-swappable units, redundancy, and bidirectional energy exchange. The ECRH system powers twelve gyrotrons (10 MW RF during startup, 1 MW during power production) using modular pulse-step high-voltage supplies with fast regulation, low fault energy, and N+1 redundancy, supported by auxiliary anode and body units. Together, these systems emphasize modularity, redundancy, and regenerative operation to deliver efficient, grid-compliant power for reliable long-pulse operation.

For more information on the electrical systems and power supplies of Helios, see the dedicated companion paper in this special issue: \cite{slepchenkov_electrical_nodate}.

\subsection{Instrumentation and control}
\label{sec:control}

The control and instrumentation systems of Helios enable safe, reliable, and continuous operation through a hierarchical architecture. At the top, the Main Control Unit (``MCU'') combines Graphics Processing Units (``GPUs'') for optimization with field-programmable gate arrays (``FPGAs'') for real-time execution, coordinating distributed controllers. Operating independently, the Safety Control System (``SCS'') uses programmable logic controllers (``PLCs'') to enforce interlocks and manage alarms. The Machine Instrumentation System collects data from diagnostics such as magnetic probes, quench-detection fibers, and plasma diagnostics including Thomson Scattering and Electron Cyclotron Emission (``ECE''). These inputs support protection, plasma control, performance optimization, and predictive maintenance. At the MCU core, the Plasma Control System (``PCS'') runs multi-rate loops on a GPU/FPGA platform, maintaining a real-time plasma model and adjusting heating, fueling, and shaping coils. Designed for deterministic timing, redundancy, and graceful degradation, the system ensures steady-state operation with high reliability and safety compliance.

For more information on the instrumentation and control of Helios, see the dedicated companion paper in this special issue: \cite{slepchenkov_control_nodate}.

\section{Conclusion}
\label{sec:conclusion}

Helios combines the natural steady-state and low recirculating power operation of the stellarator approach with simpler, programmable magnets. The design is intended to overcome the outstanding concerns around existing stellarator approaches, which include the complex coils, the required proximity of the coils to the plasma, the large overall size, and the lack of a divertor solution with efficient particle compression. 

This paper summarizes the results of high-fidelity analyses that are conducted in the companion papers in this special issue. These include nonlinear, resistive MHD evolution, nonlinear electrostatic gyrokinetic heat flux, 3D finite element structural simulations, 3D Monte Carlo neutron transport simulations, and more. The results paint a compelling picture of a uniquely practical stellarator, within the capabilities of present-day engineering, superconductors, and materials. The design is securely grounded in the last four decades of large, high-field stellarator experiments. 

In the next few years, Thea Energy plans to verify and de-risk the operation of the plasma and subsystems in its large-scale integrated stellarator facility, ``Eos'' \cite{swanson_scoping_2025}. In Eos's initial operational phase, the heating, fueling, confinement, stability, electromagnetic operation, divertor, and wall condition of a Helios-like architecture will be rigorously demonstrated. A subsequent operational phase is possible, in which beam-target deuterium-deuterium fusion further de-risks tritium breeding, neutron shielding, neutron exposure effects, tritium processing, fuel cycle, and safety features. The first plasma in Eos is targeted for 2030. The first plasma in Helios is targeted in the mid 2030s. 

\section{Acknowledgments}
\label{sec:acks}

This research was funded by Thea Energy and performed as part of the DOE Milestone-Based Fusion Development Program (DE-SC0024881).

This research used resources of the National Energy Research Scientific Computing Center (``NERSC''), a Department of Energy User Facility using NERSC award FES-ERCAP 0031504.

The simulations presented in this article were performed on computational resources managed and supported by Princeton Research Computing, a consortium of groups including the Princeton Institute for Computational Science and Engineering (``PICSciE'') and the Office of Information Technology's High Performance Computing Center and Visualization Laboratory at Princeton University.

\bibliographystyle{elsarticle-num} 
\bibliography{overview}

\end{document}